\documentclass[aps,twocolumn,showpacs,preprintnumbers,superscriptaddress]{revtex4-2}

\usepackage[utf8]{inputenc}
\usepackage[english]{babel}
\usepackage{amsmath}
\usepackage{braket}
\usepackage{bm}
\usepackage{chemformula}
\usepackage{siunitx}
\usepackage{amsbsy}

\usepackage{graphicx}
\usepackage{multirow}
\usepackage{booktabs}
\usepackage[T1]{fontenc}
\usepackage{dcolumn}  
\usepackage{float}

\usepackage{amssymb}
\usepackage{amsfonts,dsfont}
\usepackage{subfigure}

\usepackage[normalem]{ulem}
\usepackage[bbgreekl]{mathbbol}
\usepackage{relsize,scalerel}

\usepackage[linktocpage=true,
  colorlinks=true, 
  pdfborder={0 0 0},
  linkcolor=blue,
  citecolor=red,
  filecolor=yellow,
  urlcolor=blue,
  bookmarks,
  pdfauthor={},
]{hyperref}

\newcommand{\eq}{\sss{eq}}

\newcommand{\pstmmc}{\textit{I}$\bar{4}$3\textit{m}}
\newcommand{\ScH}{CH\textsubscript{4}-H\textsubscript{3}S}

\newcommand{\Rcal}{\mathcal{R}}

\newcommand{\Fcal}{\mathcal{F}}

\newcommand{\bR}{\boldsymbol{R}}

\newcommand{\bD}{\boldsymbol{D}}

\newcommand{\bq}{\boldsymbol{q}}
\newcommand{\bk}{\boldsymbol{k}}

\newcommand{\bvarPhi}{\boldsymbol{\varPhi}}

\newcommand{\bRcal}{\boldsymbol{\Rcal}}

\newcommand{\sss}[1]{\scriptscriptstyle{\text{#1}}}

\newcommand{\rscha}{\bRcal}
\newcommand{\rschatrial}{\bRcal}
\newcommand{\phischatrial}{\bvarPhi}

\newcommand{\rhoschatrial}{{\tilde\rho}_{\scriptscriptstyle{\rscha},\scriptscriptstyle{\phischatrial}}}

\newcommand{\bRcaleq}{\bRcal_{\eq}}
\newcommand{\DF}{D^{\sss{(F)}}}
\newcommand{\DHA}{D^{\sss{(H)}}}
\newcommand{\bDF}{\bD^{\sss{(F)}}}

\begin{document}

\title{Quantum Anharmonic Effects on the Superconductivity of \pstmmc\ \ScH\ at High Pressures: a First-Principles Study}

\author{Pugeng Hou}

\affiliation {College of Science, Northeast Electric Power University, Changchun Road 169, 132012, Jilin, P. R. China}

\author{Francesco Belli}

\affiliation{Department of Chemistry, State University of New York at Buffalo, Buffalo, New York 14260-3000, USA}

\author{Tiange Bi}

\affiliation {Carnegie Institution for Science, Broad Branch Road, Northwest, Washington, D.C., Washington 20015, USA}

\author{Eva Zurek}
\email{ezurek@buffalo.edu}

\affiliation{Department of Chemistry, State University of New York at Buffalo, Buffalo, New York 14260-3000, USA}

\author{Ion Errea}
\email{ion.errea@ehu.eus}

\affiliation {Fisika Aplikatua Saila, Gipuzkoako Ingeniaritza Eskola, University of the Basque Country (UPV/EHU), Europa Plaza 1, 20018 Donostia/San Sebastián, Spain} 

\affiliation {Centro de Física de Materiales (CSIC-UPV/EHU), Manuel de Lardizabal Pasealekua 5, 20018 Donostia/San Sebastián, Spain} 

\affiliation {Donostia International Physics Center (DIPC), Manuel de Lardizabal Pasealekua 4, 20018 Donostia/San Sebastián, Spain}

\date{\today}

\begin{abstract}  
Making use of first-principles calculations we analyze the effect of quantum ionic fluctuations and lattice anharmonicity on the crystal structure and superconductivity of \pstmmc\ \ScH, one of the lowest enthalpy structures in the C-S-H system, in the 150 – 300 GPa pressure range within the stochastic self-consistent harmonic approximation. We predict a correction to the crystal structure, which is formed by an H$_3$S lattice and CH$_4$ molecules, the phonon spectra, and the pressure-dependent superconducting critical temperatures, which have been estimated in previous calculations without considering ionic fluctuations on the crystal structure and assuming the harmonic approximation for the lattice dynamics. Our results show that quantum ionic fluctuations have an impact on the distance between H atoms and S atoms in the H\textsubscript{3}S host lattice, pushing it towards more symmetric bonds, while the methane molecules are barely affected. According to our anharmonic phonon spectra, this structure is dynamically stable above 150~GPa, which is 30~GPa lower than the pressure at which the harmonic approximation predicts the emergence of an instability. As a consequence of the strong anharmonic enhancement of the phonon frequencies, the electron-phonon coupling constant is suppressed by 46\% at 200~GPa, and even more at lower pressures. As a result, the superconducting critical temperature is overestimated by around 50~K at 200~GPa, such that it falls below 150~K in the whole pressure range studied. Our results underline that ternary hydrides are subject to strong anharmonic effects on their structural, vibrational, and superconducting properties.
\end{abstract}

\maketitle 

\section{Introduction}

In recent years, high-pressure hydrogen-rich compounds  have demonstrated remarkable potential in yielding a plethora of structures exhibiting high-temperature superconductivity. Experimentally, critical temperatures ($T_c$'s) exceeding 200 K have been reported notably in sulfur \cite{22}, lanthanum \cite{23,24}, yttrium \cite{25,27}, and calcium \cite{ma2022high,li2022superconductivity} hydrides under pressures above 100 GPa. Presently, the highest recorded $T_c$ is attributed to the hydrogen-rich compound LaH$_{10}$, with a value of 250 K at 150 GPa~\cite{23,24}. Despite these accomplishments, however, further work is still necessary to synthesize compounds at lower pressures while still being able to retain a  high $T_c$. In this regard, computational structural predictions based on density-functional theory (DFT) established themselves as a fundamental tool able to provide valuable insights on the existence of novel superconducting materials, while showcasing a remarkable accuracy in the predictions of their $T_c$'s. This prompts these methods to be used as a guide for experimental discovery \cite{Methods,Superconducting,dope,Zurek:2021k}, and indeed many of the synthesized compounds have been anticipated by DFT calculations~\cite{H3S,H3S-2,PH3,YH10,LaH-YH,CaH6}. 

Through these DFT-based methods, most of the binary combinations of hydrogen have now been theoretically explored. The focus has currently shifted towards the reduction of the pressure of stability by expanding the list of predicted compounds to ternary and quaternary hydrides. The few of the most prominent results are the prediction of a $T_c$ of about 450 K at 250 GPa for Li$_2$MgH$_{16}$~\cite{Li2MgH16} or 370~K at 300~GPa in a structurally related Li$_2$CaH$_{17}$ phase~\cite{Zurek:2024f}, the possible metastability down to 40 GPa of LaBH$_{8}$ in the $Fm\bar{3}m$ high-symmetry phase with a remarkable $T_c$ of approximately 120 K \cite{LaBH-liang,LaBH8}, and  LaBeH$_8$ in the $Fm\bar{3}m$ high-symmetry phase with a $T_c$ of 160 K at 20 GPa \cite{zhang2022design}, later synthesized with a $T_c$ of 110 K at 80 GPa \cite{song2023stoichiometric}. More recently there are also important predictions of metastable high-$T_c$ compounds at ambient pressure \cite{hydrides_at_ambient1}, like Mg$_2$IrH$_6$ \cite{MgIrH6_ion,MgIrH6_chris}.

Additionally, some studies have suggested the $T_c$ of H$_3$S could be enhanced through carbon or phosphorus doping \cite{dope,wang2022dilute}. However, DFT crystal structure predictions did not find any thermodynamically stable compounds in the C-S-H system at high pressures \cite{wang2021absence,CH4-H3S}, although those closest to the convex hull of stability resemble the $Im\bar{3}m$ H$_3$S structure in which one of the H$_3$S sublattices is substituted by methane molecules, resulting in a CSH$_7$ stoichiometry. Thus, if any synthesizable compound in the C-S-H system exists in the megabar regime it can probably be described as H$_3$S with encapsulated methane molecules. The different arrangements of the methane molecules and their impact on the H$_3$S sublattice give raise to many energetically competitive structures within the CSH$_7$ stoichiometry \cite{wang2021absence,CH4-H3S}. Among them the structure with \pstmmc\ space group revealed maximal $T_c$\ of 181 K at 100 GPa \cite{CH4-H3S} and 194 K at 150 GPa\cite{cui2020route}. However the impact of quantum nuclear effects and anharmonicity on doped H$_3$S systems, in particular the best CSH$_7$ candidates, and their consequent influence on structural and superconducting properties remains unexplored.

Quantum nuclear effects and anharmonicity arise when dealing with light atoms, where quantum atomic oscillations become significant and cannot be disregarded, or, similarly, when the introduction of temperature enhances the amplitude of atomic vibrations. Both cases may lead to the exploration of parts of the Born-Oppenheimer potential energy surface $V(\mathbf{R})$ that extend far beyond the limits of a second-order expansion around the local minima. These conditions define the presence of anharmonicity, driven by temperature or quantum nuclear effects. Anharmonicity  impacts across a wide spectrum of materials, ranging from charge density wave materials \cite{gutierrez2023purely,diego2021van}, perovskites\cite{ranalli2023temperature}, and most importantly a variety of phenomena in hydrogen-rich systems \cite{belli2022impact,hou2021quantum,hou2021strong,errea2020quantum,errea2016quantum,44,34,35,38,39,41,25,monacelli2019black,33,hou2021strong,ScH6}. In all these cases one is to expect renormalizations over the structural phase stability and phonon spectra, and possibly the renormalization of superconducting properties.

For hydrides these effects can both strongly enhance or suppress the predicted critical temperature. In aluminium\cite{hou2021strong,33}, palladium\cite{34,35}, and platinum\cite{35} hydrides anharmonicity hardens the H-character optical modes and suppresses superconductivity to a large degree. In the possible metallic and molecular $Cmca$ phase of hydrogen the situation is the contrary\cite{41}: anharmonicity doubles $T_c$ bringing it from around 100 K to values well above 200 K. In LaH$_{10}$\cite{44}, it has been argued that quantum effects stabilize a crystal structure with huge electron-phonon coupling that otherwise would be unstable. Also in ScH\textsubscript{6}, the enhancement of the critical temperature by approximately 15\% is due to the stretching of the H$_2$ molecular-like units \cite{ScH6}. How quantum anharmonic effects affects superconductivity is thus not known {\it a priori}.

In this work a first-principles analysis of the role of quantum ionic fluctuations and anharmonicity on the predicted high-temperature superconducting \pstmmc\ phase of \ScH\ is presented. It is shown that lattice quantum anharmonicity plays an important role also in ternary compounds. In the particular case of \pstmmc\ \ScH\  it significantly reduces the value of $T_c$. The manuscript is organized as follows: Sec. \ref{sec:methodolgy} describes the theoretical framework of our anharmonic {\it ab initio} calculations, Sec. \ref{sec:computational_details} overviews the computational details of our calculations, Sec. \ref{sec:results} discusses the results of the calculations, and Sec. \ref{sec:conclusions} summarizes the main conclusions of this work.

\section{Methodology}
\label{sec:methodolgy}

The effect of lattice quantum ionic fluctuations and anharmonicity is estimated using the stochastic self-consistent harmonic approximation (SSCHA) code\cite{monacelli2021stochastic}, whose theoretical basis was developed in Refs. \cite{34,35,51,52}. In this section we briefly review the SSCHA method  as well as the theoretical framework followed for estimating the superconducting critical temperature including anharmonicity.

\subsection{The stochastic self-consistent harmonic approximation}

The SSCHA is a quantum variational method that minimizes the 
free energy of the system calculated with a trial density matrix $\rhoschatrial$:
\begin{equation}
    \Fcal[\rhoschatrial] = \braket{K + V(\bR)}_{\rhoschatrial} - TS[\rhoschatrial].
    \label{eq:sscha_f}
\end{equation}
In the equation above, $K$ is the ionic kinetic energy, $T$ the temperature, and $S[\rhoschatrial]$ the entropy calculated with the trial
density matrix. The trial density matrix is parametrized with
centroid positions $\rschatrial$ and  auxiliary force constants $\phischatrial$. The former determine the average ionic positions and the latter are related to the broadening
of the ionic wave functions around $\rschatrial$. By minimizing $\Fcal[\rhoschatrial]$ with respect to $\rschatrial$ and $\phischatrial$, as well as calculating the stress tensor from $\Fcal[\rhoschatrial]$, the SSCHA code can optimize the crystal structure, including lattice degrees of freedom, fully including ionic quantum effects and anharmonicity at any target pressure.

Phonon frequencies within the SSCHA should not be calculated by diagonalizing the auxiliary force constants $\phischatrial$, but from the dynamical
extension of the theory\cite{51,monacelli2021timedependent,lihm2021gaussian}, which allows to determine phonon frequencies from the peaks of the single-phonon spectral function. In the static limit, the peaks coincide with the $\Omega_\mu(\bq)$ frequencies, 
where $\Omega^2_\mu(\bq)$ are the eigenvalues of the Fourier 
transform of the free energy Hessian matrix divided by the masses of the atoms:
\begin{equation}
\DF_{ab}= \frac{1}{\sqrt{M_aM_b}} 
\left[ \frac{\partial^2F}{\partial\Rcal^a\partial\Rcal^b}\right]_{\bRcaleq},
\label{eq:df}
\end{equation}
where $M_a$ is the mass of atom $a$ ($a$ and $b$ are combined indexes that determine both an atom and a Cartesian direction). 
In Eq.\ \eqref{eq:df} $F$ is assumed to be the free energy at the minimum, while $\bRcaleq$ is the centroid positions that minimize Eq.\ \eqref{eq:sscha_f}.  If $\Omega_\mu(\bq)$ is imaginary the lattice is unstable in the quantum anharmonic energy landscape, meaning that the free energy is not a minimum along the lattice distortion determined by the corresponding eigenvector.

\subsection{Calculation of the superconducting transition temperature}

We evaluate $T_c$ with the Allen–Dynes\cite{54} modified McMillan equation,
\begin{equation}
T_c = \frac{f\textsubscript{1}f\textsubscript{2}\,\omega\textsubscript{log}}{1.2} \exp \left[ -\frac{1.04(1+\lambda)}{\lambda-\mu^*(1+0.62\lambda)} \right],
\label{eq1}
\end{equation}
where $\lambda$ is the electron-phonon coupling constant and $\mu^*$ effectively parametrizes the electron Coulomb repulsion\cite{54}, $\omega_\text{log}$ is the logarithmic average phonon frequency, and $f_1$ and $f_2$ are the strong coupling and shape corrections. Despite its simplicity, this equation has yielded $T_c$ values in good agreement with experiments in hydrogen-rich compounds\cite{44}. The electron phonon coupling constant, $\lambda$, is calculated from the Eliashberg function, $\alpha^{2}F(\omega)$, as
\begin{equation}
\lambda = 2 {\int_0^\infty d\omega \frac{\alpha^{2}F(\omega)}{\omega}}.
\label{eq2}
\end{equation}
Similarly,
\begin{eqnarray}
    \omega\textsubscript{log} & = & \exp \left( \frac{2}{\lambda} \int d\omega \frac{\alpha^2F(\omega)}{\omega} \log\omega \right), \\
    f_1 & = & \left[ 1 + (\lambda / \Lambda_1)^{3/2} \right]^{1/3}, \\
    f_2 & = & 1 + \frac{(\bar{\omega}_2/\omega\textsubscript{log} - 1) \lambda^2}{\lambda^2 + \Lambda_2^2}.
\end{eqnarray}
$\Lambda_1$, $\Lambda_2$, and $\bar{\omega}_2$ are given by
\begin{eqnarray}
    \Lambda_1 & = & 2.46 (1 + 3.8\mu^*) \\
    \Lambda_2 & = & 1.82 (1 + 6.3\mu^*)(\bar{\omega}_2/\omega\textsubscript{log}) \\
    \bar{\omega}_2 & = & \left[ \frac{2}{\lambda} \int d\omega \alpha^2F(\omega) \omega \right]^{1/2}.
\end{eqnarray}

The Eliashberg function is calculated as
\begin{eqnarray}
    && \alpha^{2}F(\omega) = \frac{1}{2 N(0) N_q N_k} \sum_{\substack{\mu \bq \\ {\bk}nm \\ \bar{a}\bar{b}}}
    \frac{\epsilon_{\mu}^{\bar{a}}(\bq) \epsilon_{\mu}^{\bar{b}}(\bq)^*}{\omega_{\mu}(\bq) \sqrt{M_{\bar{a}}M_{\bar{b}}}} \nonumber \\
    && \times
    d^{\bar{a}}_{{\bk}n,{\bk}+{\bq}m} d^{\bar{b}*}_{{\bk}n,{\bk}+{\bq}m} \delta(\varepsilon_{{\bk}n})
   \delta(\varepsilon_{{\bk+\bq}m}) \delta(\omega -  \omega_{\mu} (\bq)). \nonumber \\
\label{eq:eliashberg}
\end{eqnarray}
In the equation above  $d^{\bar{a}}_{{\bk}n,{\bk}+{\bq}m} = \bra{{\bk}n}
\delta V_{KS} / \delta R^{\bar{a}}(\bq) \ket{{\bk}+{\bq}m}$, where 
$\ket{{\bk}n}$ is a Kohn-Sham state with energy $\varepsilon_{{\bk}n}$
measured from the Fermi level, $V_{KS}$ is Kohn-Sham potential, and
$R^{\bar{a}}(\bq)$ is the Fourier transformed displacement of atom
$\bar{a}$; $N_k$ and $N_q$ are the number of electron
and phonon momentum points used for the Brillouin zone sampling; $N(0)$ is the density of states 
at the Fermi level; and  $\omega_{\mu} (\bq)$ and 
$\epsilon_{\mu}^{\bar{a}}(\bq)$ represent phonon frequencies and polarization vectors.
The combined atom and Cartesian indexes with a bar ($\bar{a}$) 
only run for atoms inside the unit cell.
In this work, the Eliashberg function is calculated both at the harmonic and anharmonic levels by plugging the harmonic phonon frequencies and polarization vectors or their anharmonic counterparts obtained diagonalizing $\bDF$, respectively, into Eq. \eqref{eq:eliashberg}. It should be noted that the derivatives of the Kohn-Sham potential entering the electron-phonon matrix elements are calculated at different positions in the classical harmonic and quantum anharmonic calculations: in the former they are taken at the $\bR_0$ positions that minimize  $V(\bR)$, while in case of the latter they are taken at the $\bRcaleq$ positions that minimize $\Fcal[\rhoschatrial]$ instead. 

\section{Computational details}
\label{sec:computational_details}

All first-principles calculations were performed within DFT\cite{45} using the generalized gradient approximation (GGA) as parametrized by Perdew, Burke, and Ernzerhof (PBE)\cite{46,47}. Harmonic phonon frequencies were calculated making use of density functional perturbation theory (DFPT)\cite{DFPT_S.Baroni} as implemented in the Quantum ESPRESSO code\cite{48}. We used ultrasoft pseudopotentials, including $4$ electrons in the valence for C and $6$ electrons in the valence for S. The SSCHA\cite{monacelli2021stochastic} minimization requires the calculation of energies, forces, and stress tensors in supercells. These were also calculated with DFT at the PBE level with Quantum ESPRESSO, making use of the same pseudopotentials. The plane-wave basis cutoff for the wavefunction was set to 60 Ry and to 600 Ry for the density. Brillouin zone (BZ) integrations were performed on a 16$\times$16$\times$16 Monkhorst-Pack $\bk$-point mesh\cite{53}, using a smearing parameter of 0.02 Ry, for the unit cell harmonic phonon calculations. The SSCHA calculations were performed using a 2$\times$2$\times$2 supercell containing 288 atoms at 0 K, yielding dynamical matrices on a commensurate 2$\times$2$\times$2 $\bq$-point grid. A 50 Ry energy cutoff and a 5$\times$5$\times$5 k-point mesh for the Brillouin zone (BZ) integrations were sufficient in the supercell to converge the SSCHA gradient. Harmonic phonon frequencies and electron–phonon matrix elements were calculated on a grid of 4$\times$4$\times$4 points. The difference between the harmonic and anharmonic dynamical matrices in the 2$\times$2$\times$2 phonon-momentum grid was interpolated to a 4$\times$4$\times$4 grid. Adding the harmonic 4$\times$4$\times$4 grid dynamical matrices to the result, the anharmonic 4$\times$4$\times$4 $\bq$-grid dynamical matrices were obtained. The Eliashberg function in Eq. \eqref{eq:eliashberg} was calculated with a finer 20$\times$20$\times$20 $\bk$-point grid, with a Gaussian smearing of 0.012 Ry for the electronic Dirac deltas. 

\section{Results and discussion}
\label{sec:results}

\begin{figure*}[t]
\includegraphics[width=0.8\textwidth]{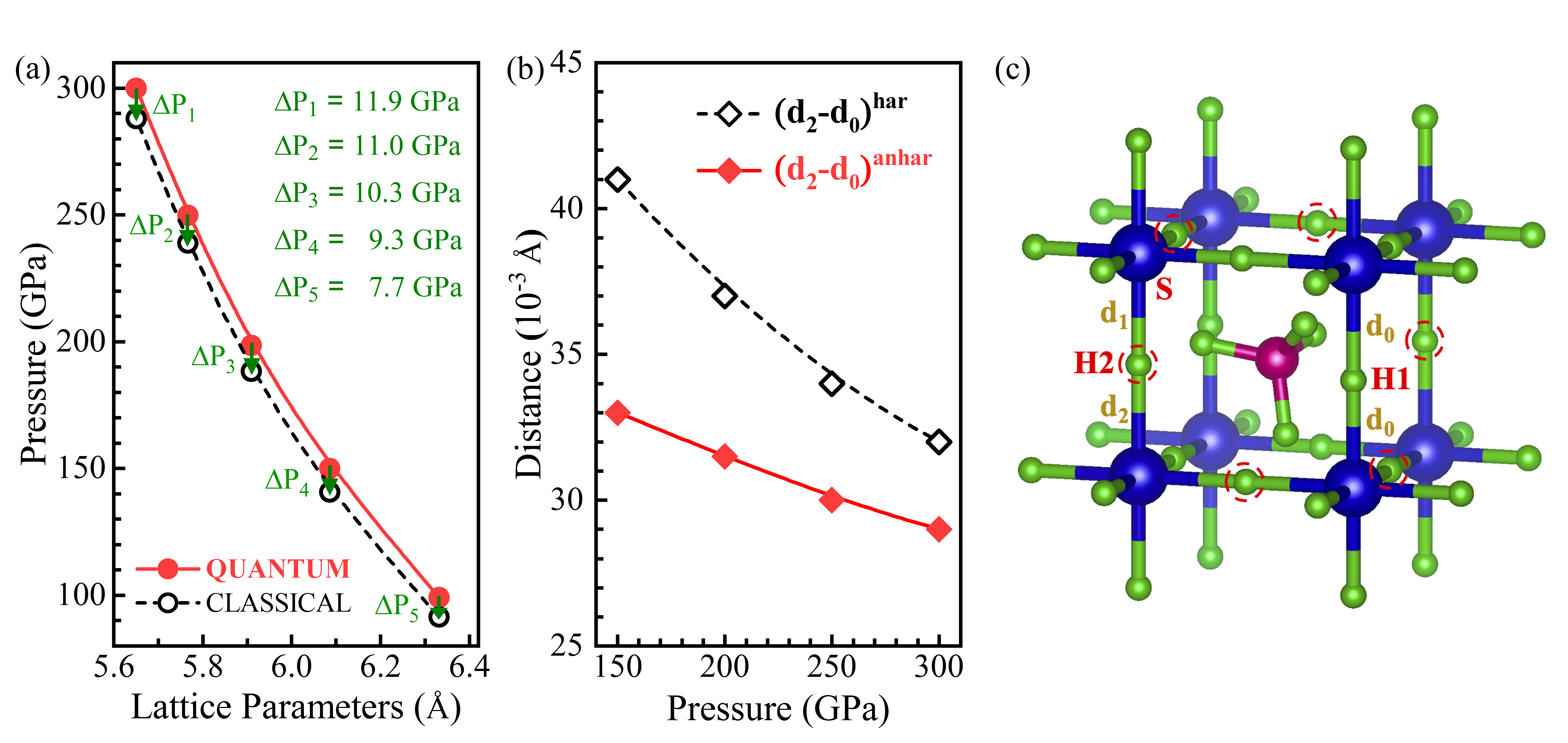}
\caption{(Color online) (a) Comparison between the classical and quantum pressures as a function of the lattice parameter of \pstmmc\ \ScH . The classical pressure is obtained from the Born Oppenheimer energy surface (BOES) and the quantum pressure from the SSCHA free energy calculations. (b) 
Difference between the distances ($d\textsubscript{2}-d\textsubscript{0}$), which measures the difference of the H2-S bond distances between the classical and quantum crystals as a function of  pressure. (c) The crystal structure of \pstmmc\ \ScH. Large blue balls represent S atoms, middle-sized purple balls represent C atoms, and small green balls represent H atoms, respectively. There are two H2-S distances denoted as $d\textsubscript{1}$ and $d\textsubscript{2}$. At all pressures we studied, $d\textsubscript{2}$>$d\textsubscript{1}$.  There are two equivalent H1-S distances marked as $d\textsubscript{0}$. For the clarity of representation, H2 atoms are marked with red circles with dashed lines.}
\label{1.jpg} 
\end{figure*}  


\subsection{Pressure and crystal structure}

The lattice of \pstmmc\ \ScH\ is a highly symmetric body-centered cubic structure with C atoms in $8c$ Wyckoff sites; S atoms in $2a$ and $6b$ sites; and H atoms in $8c$, $12d$, $12e$, and $24g$ sites \cite{CH4-H3S}. As shown in the inset in Fig. \ref{1.jpg}, the H$_3$S sublattice possesses an asymmetric H-S bond for the hydrogen labeled as H2, while a symmetric bond for the one labeled as H1. The C-H distance inside the methane molecule is not exactly the same for all bonds by symmetry, but the differences are negligible, on the order of 0.001 \AA, so that the CH$_4$ molecule can still be considered as being tetrahedral. Given the freedom in the Wyckoff positions, both the internal distances and atomic arrangements, as well as the lattice parameters could be affected by lattice quantum anharmonic effects.


As we show in Fig.\ \ref{1.jpg}(a), ionic quantum effects result in large corrections to the pressure in the equation of states. For the same lattice parameter, the pressure obtained from the classical
calculation based on the Born-Oppenheimer energy surface, or BOES (we will also refer to this as the harmonic pressure), is always about 10 GPa lower than the quantum result obtained with the SSCHA (we will also refer to this as the anharmonic pressure). This result is rather general among superhydrides between 100 and 200 GPa, as similar quantum corrections on the pressure of about 10 GPa have been estimated for LaH$_{10}$, H$_3$S, AlH$_3$\cite{39,hou2021strong,44}, and LaBH$_8$\cite{belli2022impact}.  Fig. \ref{1.jpg}(a) can be used conveniently to compare our results with previous classical calculations\cite{CH4-H3S}. For instance, we clearly mark that the classical 190 GPa and 239 GPa values correspond to 200 GPa and 250 GPa in the quantum case, respectively. In order to avoid any confusion, in the rest of the paper the pressure assigned to harmonic calculations will be the classical one, while the quantum pressure will be assigned to quantum anharmonic calculations.

As mentioned above, in the classical harmonic approximation, the atomic positions are those determined by the minimum of $V(\bR)$, the $\bR_0$ positions. If quantum anharmonic effects are considered instead, the atomic positions are determined by the minimum of the energy that includes the vibrational contribution, i.e., the zero-temperature limit of the free energy in Eq. \eqref{eq:sscha_f}. As we performed the SSCHA minimization keeping symmetries, quantum anharmonic effects optimize precisely the free parameters in the crystal structure. Our results show that quantum anharmonic effects significantly modify the position of the H2 atoms. The most important effect is that the hydrogen atoms (H2) move towards the mid-point of two nearest S atoms (see Fig. \ref{1.jpg}(b)). Thus, the H2 atoms in the quantum crystal of \pstmmc\ \ScH\ are much closer to the middle position of two S atoms than they are in the harmonic case, resembling the hydrogen symmetrization occurring in H$_3$S \cite{errea2016quantum}. It is also noteworthy that quantum anharmonic effects do not change  the guest methane lattice and the position of the H1 hydrogen atoms, which are already at the mid-point. At all pressures we considered, quantum anharmonic effects impose a similar change in the structure. Interestingly, these effects become less important as the pressure increases. This suggests that the impact of anharmonicity is suppressed in this system with pressure, as has been observed for its parent structure, $Im\bar{3}m$ H$_3$S \cite{errea2016quantum,38}. 



\begin{figure*}[t]
\includegraphics[width=1.0\textwidth]{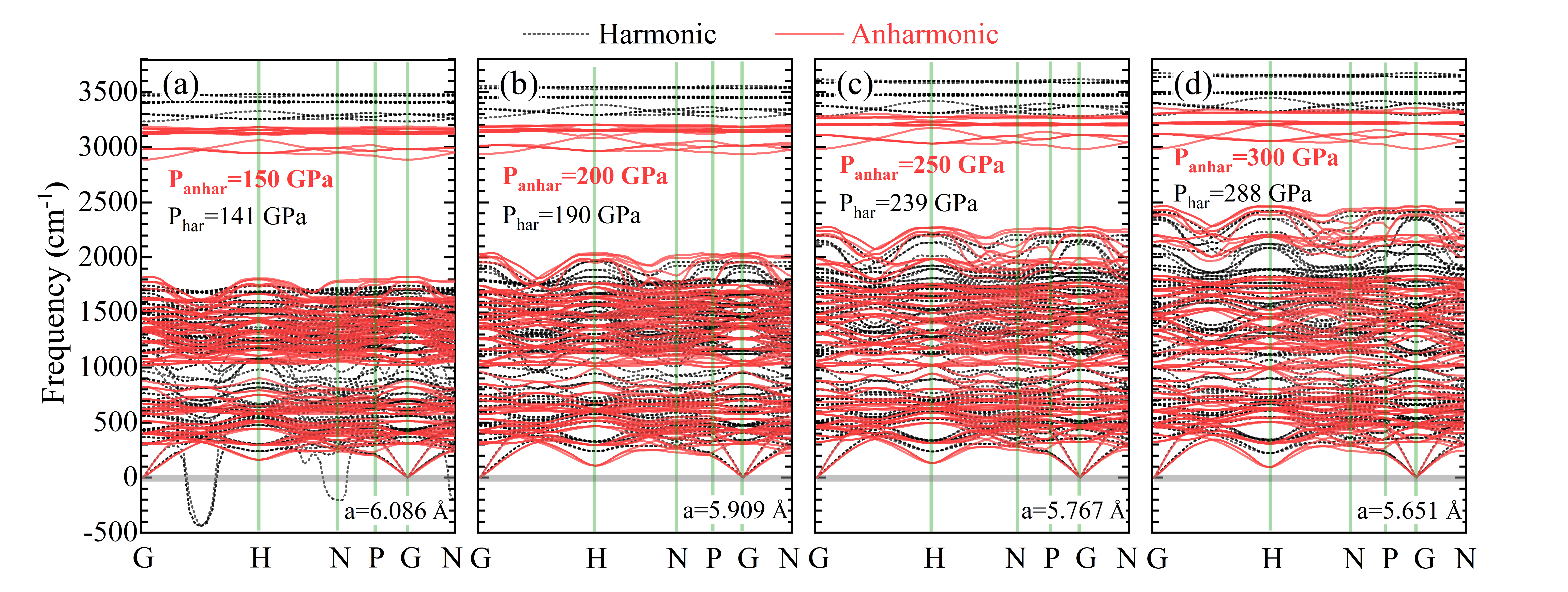}
\caption{(Color online) Comparison between the harmonic (black dot lines) and anharmonic (red solid lines) phonon spectra of the cubic high-symmetry \pstmmc\ phase of \ScH\ for different lattice parameters. The anharmonic spectra are obtained from $\bDF$ and correspond to the static limit of the SSCHA dynamical theory.
The pressure calculated classically (harmonic calculation) 
and with quantum effects (anharmonic calculation) is marked in each case. The region of positive and negative (representing imaginary) frequencies, are separated with a thick  gray line.}
\label{2.jpg} 
\end{figure*}


\subsection{Phonon Spectrum}

Quantum anharmonic effects not only affect the structure, but also the phonon spectra. Fig. \ref{2.jpg} compares the harmonic phonons obtained diagonalizing the harmonic dynamical matrix, 
\begin{equation}
\DHA_{ab} = \frac{1}{\sqrt{M_a M_b}} 
\left[ \frac{\partial^2V(\bR)}{\partial R^a \partial R^b} \right]_{\bR_0},
\label{eq:df_h}
\end{equation}
with the anharmonic ones obtained from Eq. \eqref{eq:df} at quantum pressures: (a) 150, (b) 200, (c) 250 and (d) 300 GPa, respectively. The anharmonic correction of the phonon spectra given by the SSCHA is huge. As shown in Fig. \ref{2.jpg}, approximately below 141 GPa the system develops phonon instabilities at the N point of the BZ in the classical harmonic calculation. On the contrary, the anharmonic phonons obtained diagonalizing $\bDF$ are always stable.
Therefore, quantum anharmonic effects play a crucial role in stabilizing the \pstmmc\ phase of \ScH\ around 150 GPa. The SSCHA renormalization of the phonons clearly indicates that the anharmonic correction leads to strong changes in the spectrum both for the low-energy and high-energy optical modes. The phonon spectra can be separated into three different regions: below 1000~cm$^{-1}$, where molecular rotations of the CH$_4$ are present together with modes with sulfur character related to the H$_3$S sublattice; the region between 1000 and 2000~cm$^{-1}$, predominantly of H character associated with the the H$_3$S sublattice; and the region around 3000 ~cm$^{-1}$, which present stretching modes of the CH$_4$ molecule. As expected for stretching modes, anharmonicity softens them. The behavior for the CH$_4$ rotational modes and those related to the H$_3$S sublattice is not that clear, as there are modes that are softened by anharmonicity while others become hardened.


As shown in Tab. \ref{tab:my-table}, we calculated the average phonon frequency, $\left\langle\omega\right\rangle$, both at the harmonic and anharmonic level. The value of the average phonon frequency, obtained as
\begin{equation}
\left\langle\omega\right\rangle = \frac{\int_0^\infty g(\omega) \omega d\omega}{\int_0^\infty g(\omega) d\omega},
\label{eq8}
\end{equation}
where $g(\omega)$ is the phonon density of states (PDOS), reflects that on average anharmonicity softens phonons. Our results give $(\left\langle\omega\right\rangle\textsubscript{harmonic} /\left\langle\omega\right\rangle\textsubscript{anharmonic})^2$ = 1.12 at 200 GPa. This ratio is similar at  higher pressures. At 150 GPa, this ratio reaches a smaller value of 1.09, reflecting the importance of the hardening of the modes, and in line with the fact that this structure is stabilized by anharmonicity at this pressure.

\begin{table}
\caption{Calculated average phonon frequencies of \pstmmc\ \ScH\ calculated within the classical harmonic approximation and at the quantum anharmonic level at quantum pressures of 150, 200, 250 and 300~GPa.}
\label{tab:my-table}
\centering
\begin{tabular}{cccccc}
\hline
\hline
Pressure (GPa)     & $150$    & $200$    & $250$     & $300$  \\ 
\hline 
Frequency-Har (cm$^{-1}$)   & 1390    & 1472    & 1545     & 1590 \\   
Frequency-Anh (cm$^{-1}$)   & 1330    & 1394    & 1459     & 1502  \\
\hline
\hline
\end{tabular}
\end{table}


\begin{figure}[t]
\includegraphics[width=1.0\columnwidth]{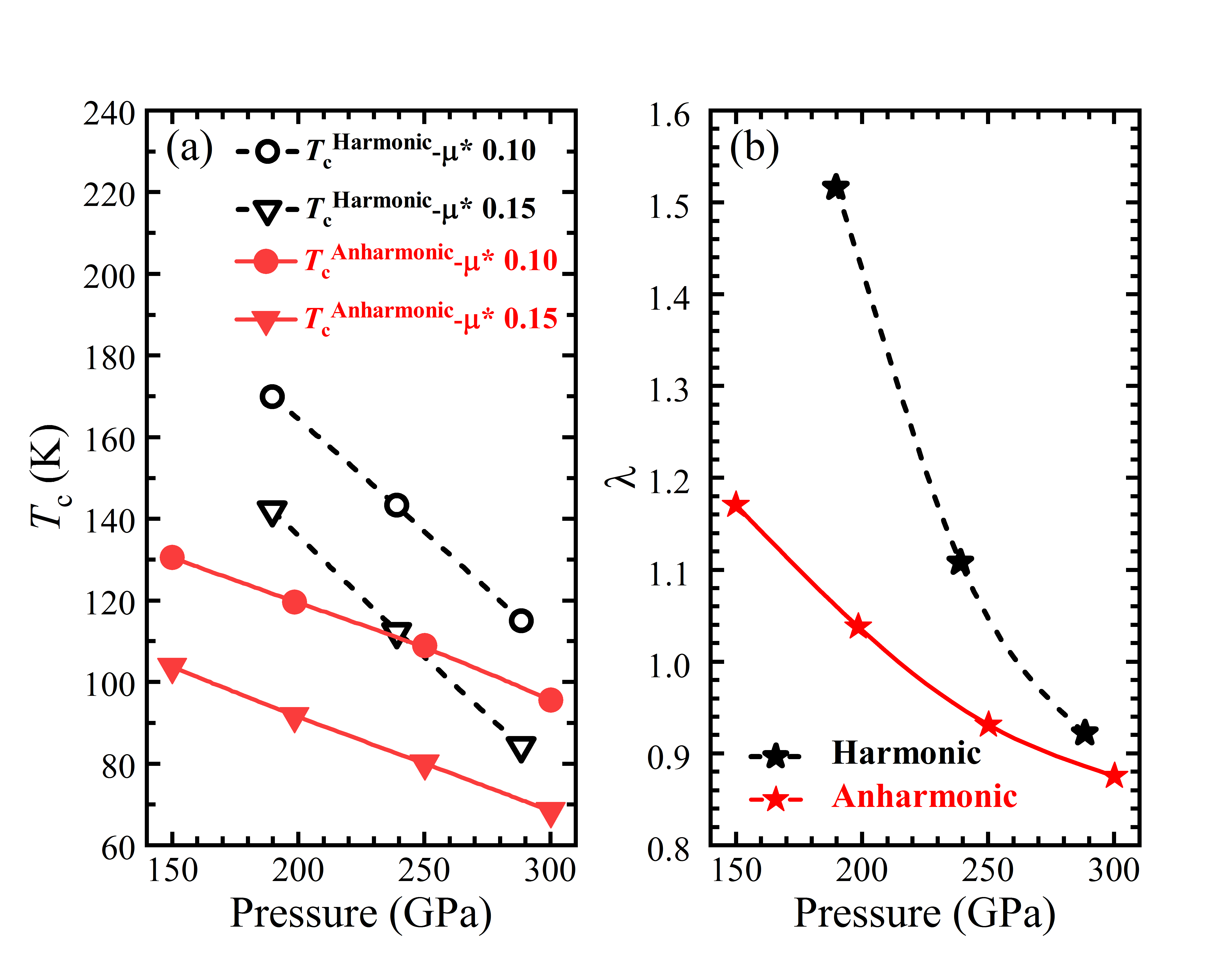}
\caption{(Color online) Superconducting critical temperature, $T_c$, (a) and electron-phonon coupling constant, $\lambda$, (b) as a function of pressure in the harmonic approximation (black dotted lines) and considering anharmonic effects (red solid lines). 
Note that harmonic and anharmonic results, which are calculated with the same lattice parameter, have the different pressures.
$T_c$ calculated with $\mu^*$ = 0.10 and 0.15 is plotted with circles and triangles, respectively. }
\label{3.jpg} 
\end{figure}


To further assess the kinetic stability of the \pstmmc\ phase of \ScH\ at lower pressures, where molecular , we performed ab-initio molecular dynamics (AIMD) simulations at 300~K and 120~GPa (see supplementary material) in order to overcome the inherent limitations of the SSCHA method with rotational degrees of freedom ~\cite{Siciliano2024Beyond}. The compound was found to be stable at this pressure, indicating that its stability extends from the previously suggested 180 GPa (based on harmonic calculations) down to at least 120 GPa at high temperatures, in agreement with the SSCHA calculations that reduce the stability pressure too. This suggests that the structure can be obtained at pressures at least 60 GPa lower than expected from classical harmonic calculations. Interestingly, during the dynamical trajectory it was observed that the CH$_4$ molecule is able to change orientation inside the H$_3$S cage overtime. This suggests that the CH$_4$ unit behaves as a rotor inside the H$_3$S cage at room temperature and 120 GPa. 
 
\subsection{Superconductivity}

Once the renormalized anharmonic phonon spectra has been obtained using the SSCHA, anharmonic effects can be easily incorporated into the electron-phonon coupling calculations as explained in Sec.\ \ref{sec:methodolgy}. We calculate $T_c$ with typical values of $\mu^*$, such as 0.10 and 0.15 (Fig.\ \ref{3.jpg}). For a given value of $\mu^*$, in the anharmonic calculation $T_c$ approaches a linear slow decrease with increasing pressure in the 150 - 300 GPa range. On the contrary, there is a clear rapid decreasing trend at the harmonic level, the same behavior described in previous works\cite{CH4-H3S}. Due to the instabilities present in the harmonic phonon spectra below 150 GPa, $T_c$ cannot be calculated below this pressure in the harmonic case for \pstmmc\ \ScH. The fact that $T_c$ soars with pressure lowering is in accordance with the general goal of obtaining materials with high critical temperature at lower pressures. The EPC constant, $\lambda$, follows the same trend under pressure as $T_c$, suggesting that the evolution of $T_c$ is governed by $\lambda$. The superconducting behavior of \pstmmc\ \ScH\ changes radically when the anharmonic renormalization of the atomic positions and phonons is considered. As shown in Fig.\ \ref{3.jpg}, anharmonicity makes the electron-phonon coupling constant drop down to a value of 1.04 at 200 GPa and 0.93 at 250 GPa, a value that is 32\% smaller than the harmonic one at this pressure. This huge anharmonic weakening of the EPC brings $T_c$ down to a value about 50~K lower than the harmonic one. A similar correction is observed at other pressures. It is interesting to remark that, despite the fact that anharmonicity on average softens the phonons, at the same time it suppresses $\lambda$ indicating that the modes that are hardened by the anharmonic renormalization are those that couple more strongly to the electrons. In fact, $\lambda$ is inversely proportional to the square of the phonon frequency. 

\begin{figure}[t]
\includegraphics[width=1.0\columnwidth]{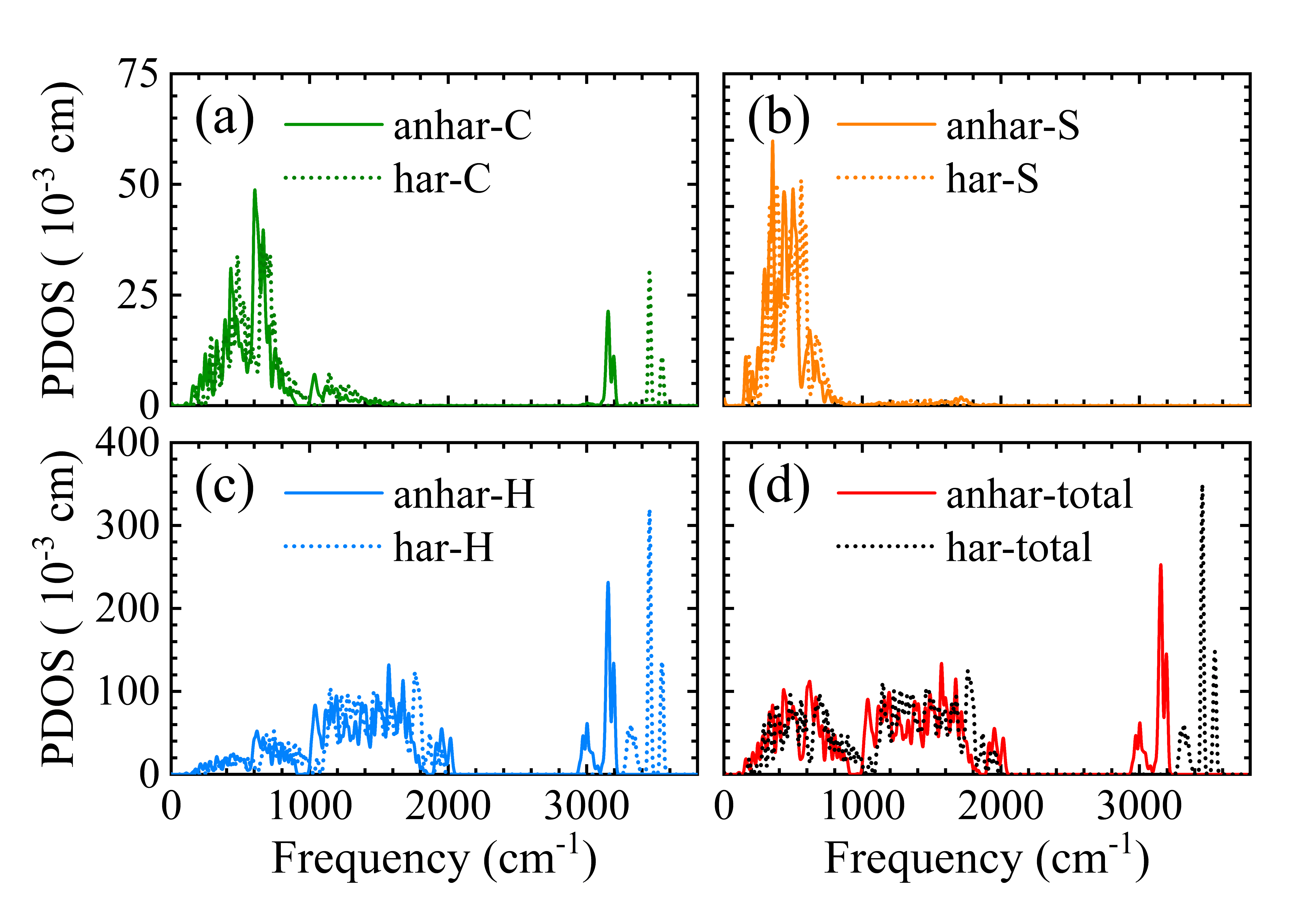}
\caption{(Color online) Projected phonon density of states (PDOS) of \pstmmc\ \ScH. Contribution from (a) C, (b) S, (c) H atoms and (d) total PDOS. For comparison, the harmonic results are also shown with the same lattice parameter 5.909 a$_0$, which corresponds in the quantum anharmonic case to 200 GPa. The harmonic results are shown with green (a), orange (b), blue (c) and black (d) dot lines, and the anharmonic results are shown with the same color in (a), (b), (c) and red in (d) with solid lines, respectively.}
\label{4.jpg} 
\end{figure}

\begin{figure}[t]
\includegraphics[width=1.05\columnwidth]{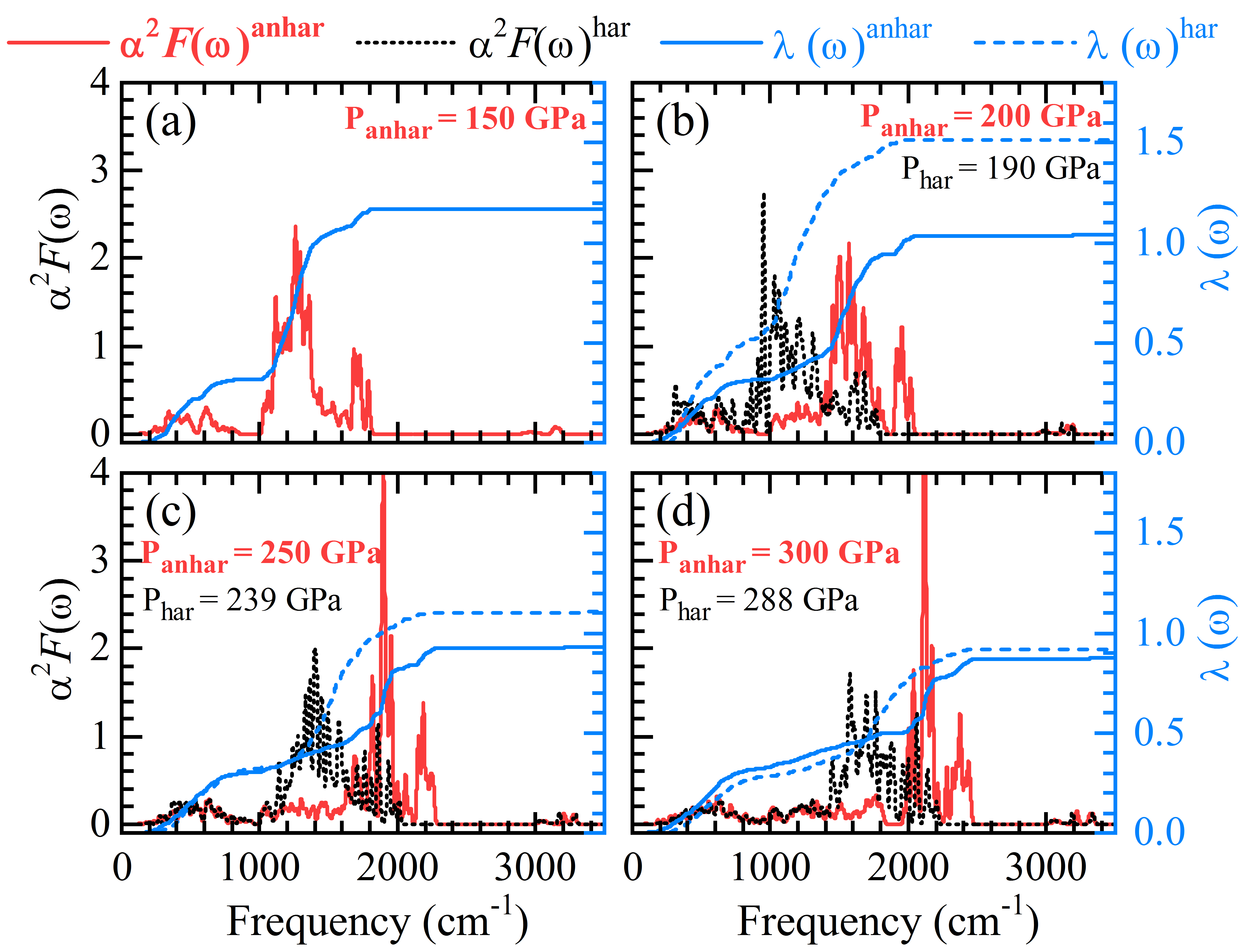}
\caption{(Color online) Anharmonic Eliashberg spectral function, $\alpha^2F(\omega)$, (red solid lines) and integrated
electron-phonon coupling constant $\lambda(\omega)$ (blue solid lines) 
at four different quantum pressures: (a) 150 GPa, (b) 200 GPa, (c) 250 GPa,
and (d) 300 GPa, respectively. The harmonic spectral function and integrated
electron-phonon coupling constant
are also shown with black and blue dotted lines for comparison at three different classical pressures: (b) 190 GPa, (c) 239 GPa and (d) 288 GPa.
The harmonic and anharmonic results in each panel are obtained with the same 
lattice parameter.}
\label{5.jpg} 
\end{figure}

In order to get a deeper understanding of the origin of such an anharmonic effect, it is convenient to have a look into the projected phonon density of states (PDOS) (Fig.\ \ref{4.jpg}), $\alpha^{2}F(\omega)$, and the integrated electron-phonon coupling constant (Fig.\ \ref{5.jpg})
\begin{equation}
    \lambda(\omega) = 2 {\int_0^\omega d\Omega \frac{\alpha^{2}F(\Omega)}{\Omega}}.
\end{equation}
The contribution to the PDOS from C, S and H atoms is clearly separated. Almost  all the contribution in the mid-high frequency region (beyond 1000 cm$^{-1}$) comes from the H atoms, while the low-frequency vibrations (below 1000~cm$^{-1}$) are dominated by the heavy S and C atoms. As expected in these types of superconductors, the mid-high frequency region (1000 - 1500 and 2000 - 2500~cm$^{-1}$) is due to vibrations of the H atoms, having the largest contribution to $\lambda$, around 70\% of the total. As shown in Fig. \ref{5.jpg}(b), the electron-phonon coupling constant is weakened in the anharmonic case mostly because of the phonon hardening induced by anharmonicity, which blueshifts the Eliashberg function. 

\section{Conclusions}
\label{sec:conclusions}

In summary, this work demonstrates the importance of quantum effects and the consequent anharmonicity on the structural and superconducting properties of \pstmmc\ \ScH\ under high pressure. Quantum anharmonic effects affect the position of half of the H atoms in the H{\textsubscript{3}}S host lattice. Consequently, the phonon spectra are strongly affected. Quantum anharmonic effects make the structure dynamically stable at least down to 150 GPa, which is at least 30 GPa lower than expected classically within the harmonic approximation, while AIMD simulations at 300~K suggest that the structure is also stable down to 120~GPa, even if the CH$_4$ acquires a rotor behavior. Anharmonicity decreases the electron-phonon coupling by 46\% at 200 GPa, and even more at lower pressure. The suppression in $\lambda$ results in an overestimation of $T_c$ by around 50~K at 200~GPa in the harmonic case. We determine that the decrease is directly a consequence of the hardening the part of the phonon spectrum that couples more strongly to the electrons. 

\section*{Acknowledgements}

P.H. thankfully acknowledges the International Science and Technology Cooperation Project of Jilin Provincial Department of Science and Technology (Grant No. 20240402055GH). E.Z. acknowledges the National Science Foundation (Grant No.
DMR-2136038) for financial support. I.E. acknowledges funding from ERC under the European Unions Horizon 2020 research and innovation program (Grant Agreements No. 802533); the Department of Education, Universities and Research of the Eusko Jaurlaritza, and the University of the Basque Country UPV/EHU (Grant No. IT1527-22); and the Spanish Ministerio de Ciencia e Innovación (Grant No. PID2022-142861NA-I00).

\bibliography{biblio}
\bibliographystyle{apsrev4-2}

\end{document}


\maketitle

\section{\emph{Ab Initio} Molecular Dynamics Simulation}
An \emph{ab initio} molecular dynamics (AIMD) simulation was performed for $I\bar{4}3m$ \ScH\ at 300~K and 120~GPa using the Vienna \emph{ab-initio} Simulation Package (VASP) version 5.4.1 \cite{Kresse:1993a, Kresse:1999a}. PAW pseudopotentials were used within the Perdew-Burke-Ernzerhof (PBE)~\cite{GGA-PBE} parametrization of the exchange-correlation functional. The C 2s$^{2}$2p$^{6}$3s$^{0}$, S 3s$^{2}$3p$^{4}$3d$^{0}$ and H 1s$^{1}$2p$^0$, electronic configurations were treated explicitly. The simulation was performed using a $2\times2\times4$ supercell with a wavefunction cutoff of 600~eV, and the $k$-mesh including only a single point at $\Gamma$.

\begin{figure}
    \centering
    \includegraphics[width=0.4\linewidth]{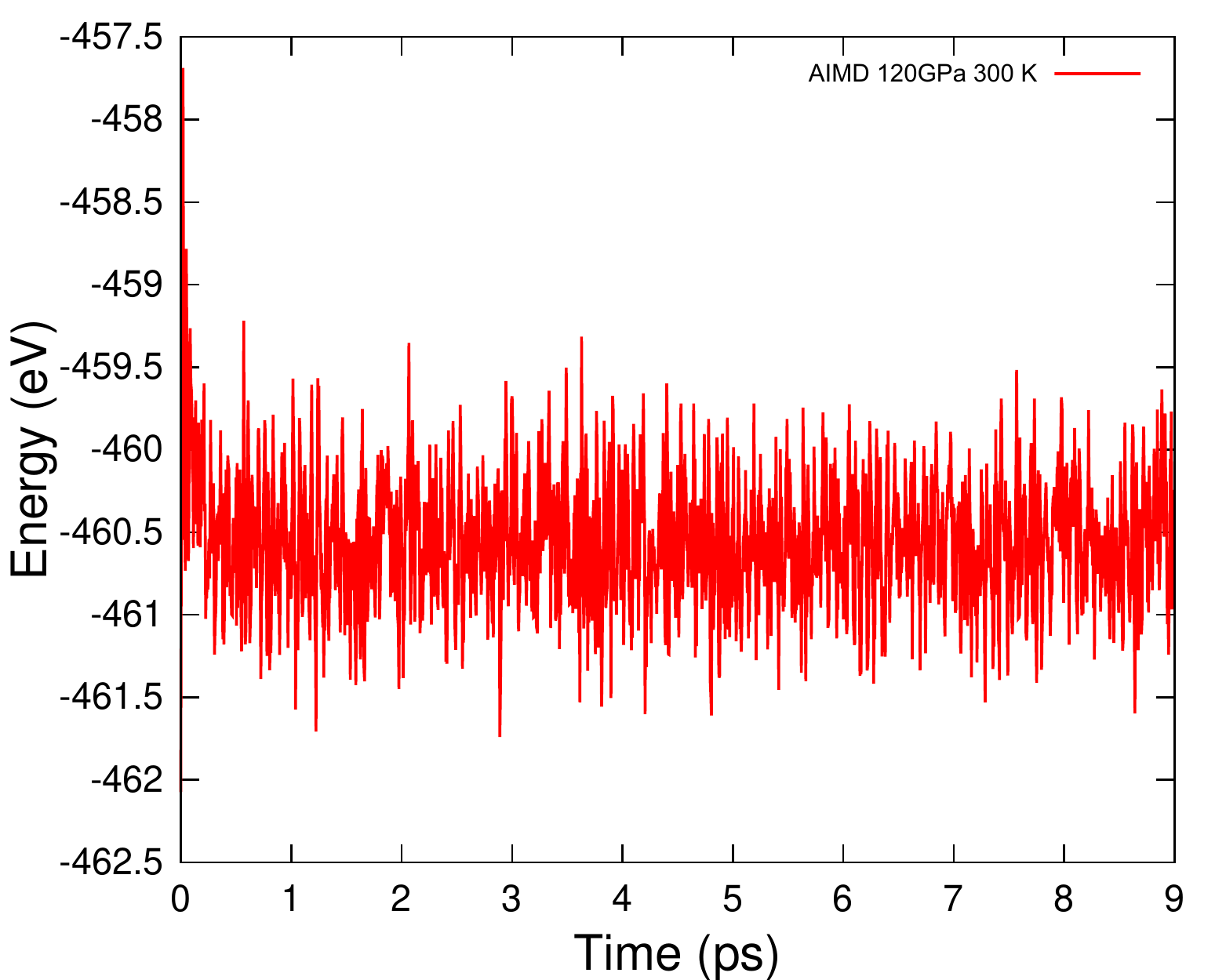}
    \includegraphics[width=0.4\linewidth]{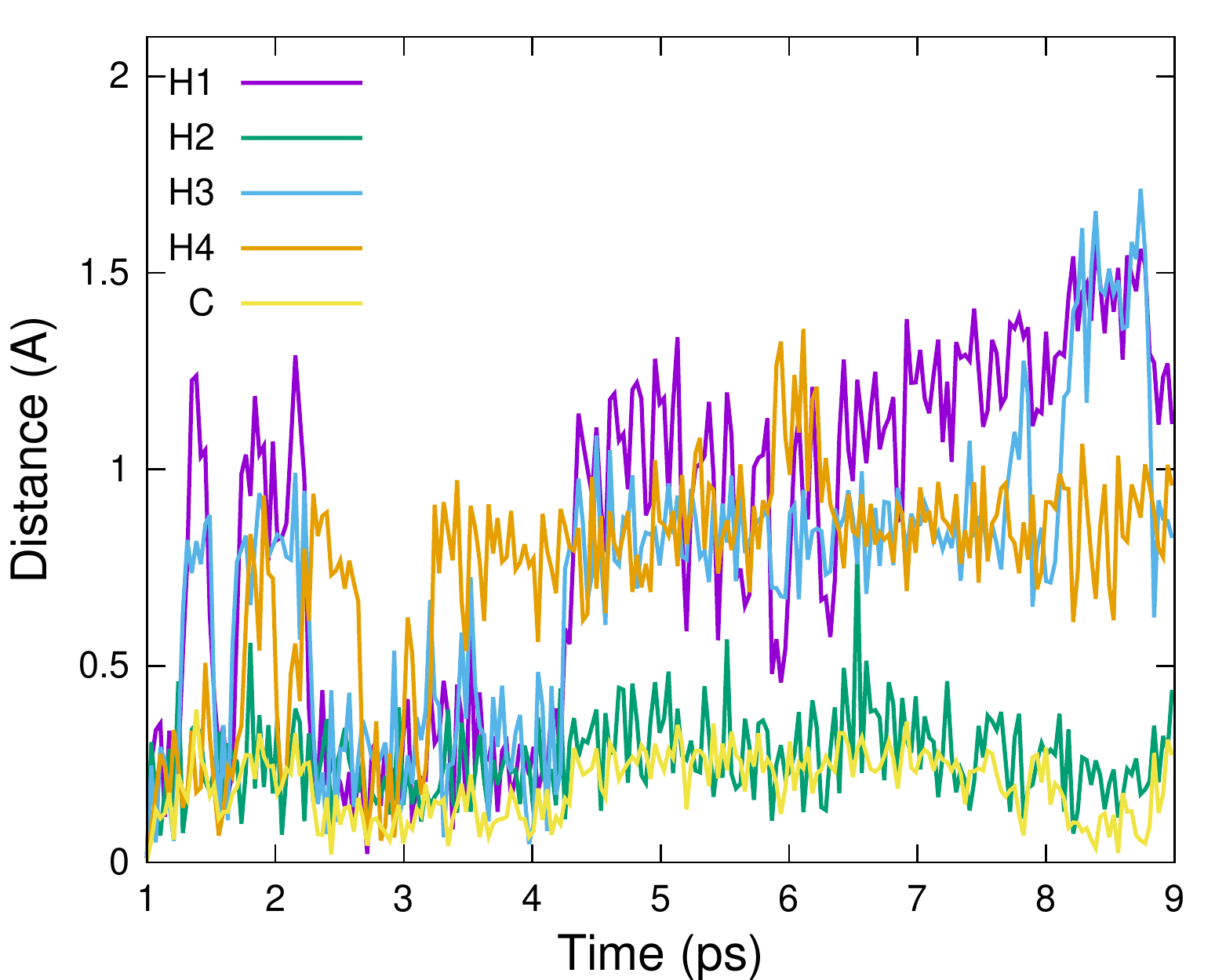} 
    \caption{(left) The temporal evolution of energy of the system during the AIMD simulation, (right) alongside the atomic displacements of each atom in a CH$_4$ molecule. The system reaches thermal equilibrium at approximately 1 ps, however after that time the CH$_4$ molecule appears to undergo rotational motion within the H$_3$S cage along one of its C-H axes.}
    \label{fig:sup1}
\end{figure}

A plot of the energy as a function of time is reported in Figure \ref{fig:sup1} (left). The AIMD simulation was performed for 9~ps. After a short equilibration, the energy oscillates about an average value, highlighting that at 120~GPa and 300~K the compound is found to be kinetically or thermally stable. More interestingly, during the dynamics it was observed that the CH$_4$ units do not simply undergo bond-stretching vibrations, but that overtime they are able to change orientation inside the H$_3$S cage by rotating along one of the C-H axes of the molecule. This is highlighted by a large variation of only three out of four of the C-H distances as a function of time for the hydrogen atoms, which can be observed by plotting the displacement of the atoms of the CH$_4$ from a reference position chosen randomly during the dynamics (Figure \ref{fig:sup1} (right)). This suggests that the CH$_4$ units behave as a rotor inside the H$_3$S cage, at least at room temperature. In this case the reference position for each atom was chosen from a snapshot at 1 ps to ensure the system achieved thermal equilibrium.

\bibliography{bib_sup}